%% file: H2889.tex
\documentclass{aa}
\usepackage{natbib}
\input{rrmacros2e.tex}
\usepackage{graphics}
\bibpunct{(}{)}{,}{a}{}{,}

\begin{document}
\title{The influence of dynamic tides on the apsidal-motion rate in
  close binaries with an evolved main-sequence star} 
\author{B.\ Willems\inst{1} \and A. Claret\inst{2}}
\institute{Department of Physics and Astronomy, The Open University,
Walton Hall, Milton Keynes, MK7 6AA, UK \\
\email{B.Willems@open.ac.uk}
\and
Instituto de Astrof\'{\i}sica de Andaluc\'{\i}a, CSIC, Apartado 3004, 
E-18080 Granada, Spain \\
\email{claret@iaa.es} }
\date{Received date; accepted date} 

\abstract{The validity of the classical formula for
  the rate of secular apsidal motion in close binaries is investigated
  for a sequence of models of a $5\,M_\odot$ star ranging from the
  last stages of the ${\rm C}^{12} \rightarrow {\rm N}^{14}$ reaction
  to the phase where 
  hydrogen is exhausted in the core. For binaries with short orbital
  periods, the apsidal-motion rates
  predicted by the classical formula deviate from the rates determined
  within the framework of the theory of dynamic tides due to the
  effects of the compressibility of the stellar fluid and due to
  resonances of dynamic tides with free oscillation modes of the
  component stars \citep{SW2001}. As the star evolves on the main
  sequence, the deviations caused by the compressibility of the
  stellar fluid increase with increasing radius of the star. The
  additional deviations caused by the resonances are largest near the
  end of the core-hydrogen burning phase. Both of these deviations
  increase with increasing values of the orbital eccentricity. 
}

\titlerunning{Apsidal motion in binaries with an evolved main-sequence star}

\maketitle

\section{Introduction}

In close binary systems of stars, each star experiences the
time-dependent tidal force exerted by its companion. The distortion
caused by the tidal force leads to a perturbation of the external
gravitational field, which in its turn causes secular variations of
the position of the periastron in the relative orbit of the companion.

The most commonly used formula for the rate of secular apsidal motion
due to the tidal deformation of the components of a close binary was
derived by \citet{Cow1938} and subsequently by
\citet{Ste1939} under the assumption that the orbital period
is long in comparison with the free harmonic periods of the component
stars. A generalisation of this classical formula taking into account the
effects of an inclination of the equatorial planes of the stars with
respect to the orbital plane of the binary was derived by
\citet{Kop1959}.  

In a recent investigation, \citet{SW2001} (hereafter referred to as
Paper~ I) reconsidered the derivation
of the classical apsidal-motion formula in the
light of the theory of dynamic tides. The authors concluded that, for
sufficiently long orbital periods, the tide raised by the companion can
be approximated {\it at each instant} by an appropriate linear
combination of three static tides. For binaries with sufficiently long
orbital periods, the classical formula then turns out to be valid up to
high orbital eccentricities. When the orbital period is shorter,
deviations arise due to the larger role of the stellar compressibility
at higher forcing frequencies and due to the effects of resonances
between dynamic tides and free oscillation modes. From the application
of their results to $5\,M_\odot$, $10\,M_\odot$, and $20\,M_\odot$
zero-age main sequence stars, \citeauthor{SW2001} found the relative
differences between the classical formula and the formula established
within the framework of the theory of dynamic tides to be inversely
proportional to the square of the orbital period. The relative
differences also turned out to be larger for a model with a larger
mass.  

In this paper, we investigate the validity of the classical formula 
for the rate of secular apsidal motion in binaries with an evolved
main-sequence star. Deviations between the classical formula 
and the formula established in the framework of the theory of dynamic
tides may be expected to vary with the evolutionary stage of the star
due to the changes in the central condensation and due to the growth 
of the stellar radius.  

The plan of the paper is as follows. In Sects.~2 and~3, we present the
basic assumptions and the ingredients necessary for the determination
of the rate of secular apsidal motion in binaries with eccentric orbits.  
In Sect.~4, we briefly recall the classical formula for the rate
of secular 
apsidal motion and the corresponding formula established within the
framework of the theory of dynamic tides.  
In Sect.~5, we examine the influence of stellar evolution on the
deviations between the apsidal-motion rates predicted by the classical 
formula and the rates given by the theory of dynamic tides for
binaries with shorter orbital periods. Sect.~6 is devoted to
concluding remarks.

\section{Basic assumptions}

Consider a close binary system of stars that are orbiting around each
other in an unvarying Keplerian orbit with semi-major axis $a$ and
orbital eccentricity $e$. The first star, with mass $M_1$, is rotating
uniformly around an axis perpendicular to the orbital plane with an angular
velocity $\Omega$ which is assumed to be small in comparison to the
inverse of the star's dynamical time scale:  
\begin{equation}
\Omega \ll {1 \over \tau_{\rm dyn}} \equiv \left( {{G\, M_1} \over R_1^3}
  \right)^{1/2}.  \label{omega}
\end{equation}
Here $G$ is the Newtonian constant of gravitation and $R_1$ the
radius of the undistorted spherically symmetric equilibrium star. We
treat the companion star, with mass $M_2$, as a point mass. 

Let $C_1xyz$ be an orthogonal right-handed frame of reference whose origin
coincides with the mass centre $C_1$ of the tidally distorted star. The
$z$-axis is normal to the orbital plane and is oriented in the sense
of the binary's orbital angular momentum. The $x$- and $y$-axes are
assumed to be  corotating with the star.  
With respect to this frame of reference, we introduce a system of
spherical coordinates $\vec{r} = (r,\theta,\phi)$ so that at each
instant $t$ the position of the companion is given by   
\begin{equation}
r=u, \;\;\; \theta=\pi/2, \;\;\; \phi=v-\Omega\,t,  \label{co}
\end{equation}
where $u$ is the radial distance and $v$ the true anomaly of the
companion in its relative orbit. 

The tidal force exerted by the companion is derived from the
tide-generating potential $\varepsilon_T\, W \left( \vec{r},t
\right)$, where $\varepsilon_T$ is a small dimensionless parameter
defined as 
\begin{equation}
\varepsilon_T = \left( {R_1 \over a} \right)^3\, {M_2 \over M_1}.
  \label{epsT}
\end{equation}
The parameter corresponds to the ratio of the tidal force to the
gravity at the star's equator.  

Following \citet{Pol1990}, we decompose the tide-gener\-a\-ting
potential in terms of unnormalised spherical harmonics
$Y_\ell^m(\theta,\phi)$ and in Fourier series in terms of multiples of
the companion's mean motion $n=2\pi/P_{\rm orb}$, where $P_{\rm orb}$
is the orbital period. The decomposition takes the form  
\begin{eqnarray}
\lefteqn{\varepsilon_T\, W \left( \vec{r},t \right) = -
  \varepsilon_T\, {{G\, M_1} \over R_1}\,
  \sum_{\ell=2}^4 \sum_{m=-\ell}^\ell \sum_{k=-\infty}^\infty}
  \nonumber \\
& & c_{\ell,m,k}\, \left( {r \over R_1} \right)^\ell
  Y_\ell^m (\theta,\phi)\, \exp \left[ {\rm i}
  \left( \sigma_T\, t - k\, n\, \tau \right) \right],
  \label{pot}
\end{eqnarray}
where $\sigma_T = k\, n + m\, \Omega$ is the forcing angular frequency
with respect to the corotating frame of reference, $\tau$ is a time of
periastron passage, and the $c_{\ell,m,k}$ are Fourier coefficients
given by  
\begin{eqnarray}
\lefteqn{c_{\ell,m,k} = \displaystyle
  {{(\ell-|m|)!} \over {(\ell+|m|)!}}\, P_\ell^{|m|}(0)
  \left({R_1\over a}\right)^{\ell-2}
  {1\over {\left({1 - e^2}\right)^{\ell - 1/2}}} } \nonumber \\
 & & {1\over \pi} {\int_0^\pi (1 + e\, \cos v)^{\ell-1}\,
  \cos (k\, M + m\, v)\, dv}. \label{pot:2}
\end{eqnarray}
In the latter expression, $P_\ell^{|m|}(x)$ is an associated Legendre
polynomial of the first kind, and $M$ is the mean anomaly of the
companion.  

The Fourier coefficients $c_{\ell,m,k}$ are equal to zero for odd
values of $\ell+|m|$ and decrease with increasing values of the
multiple $k$ of the companion's mean motion. The decrease is slower
for higher orbital eccentricities, so that the number of terms that
has to be taken into account in the expansion of the
tide-generating potential given by Eq.~(\ref{pot}) increases with 
increasing values of the orbital eccentricity. 

It follows that, for each degree $\ell$ of the spherical harmonics $Y_\ell^m
(\theta,\phi)$, the tide-generating potential consists of a term
giving rise to a static tide $(\sigma_T = 0)$, and an infinite number
of terms giving rise to partial dynamic tides $(\sigma_T \ne 0)$. 
The static tides are not to be assimilated with equilibrium tides
which occur when the orbit of the companion is circular and the
rotation of the star is synchronised with the orbital motion of the
companion.

\section{The equations governing forced isentropic oscillations of a 
spherically symmetric star}

When the effects of the Coriolis force and the centrifugal force are
neglected, the uniformly rotating star can be considered to be 
spherically symmetric. In the isentropic approximation, a partial
dynamic tide generated by a single term in the expansion of 
the tide-generating potential is then governed by the homogeneous
fourth-order system of differential equations  
\begin{eqnarray}
\lefteqn{{{d \left( r^2 \xi_T \right)} \over {dr}} =
  {g \over c^2}\, r^2 \xi_T
  + \left[ \ell(\ell+1)
  - {r^2 \over c^2}\, \sigma_T^2 \right] \eta_T
  + {r^2 \over c^2}\, \Psi_T,} \label{dyn:1} \\
\lefteqn{{{d \eta_T} \over {dr}} = \left( 1 - {N^2 \over
\sigma_T^2}
  \right) \xi_T + {N^2 \over g}\, \eta_T - {1 \over \sigma_T^2}\,
  {N^2 \over g}\, \Psi_T,} \label{dyn:2} \\
\lefteqn{{1 \over r^2}\, { d \over {dr}} \left( r^2
  {{d\Psi_T} \over {dr}} \right) - {{\ell(\ell+1)} \over r^2}\,
  \Psi_T  } \nonumber \\
 & & = 4\, \pi\, G\, \rho\, \left[ {N^2 \over g}\, \xi_T + {1 \over c^2}
  \left( \sigma_T^2\, \eta_T - \Psi_T \right) \right].
  \label{dyn:3}
\end{eqnarray}
In these equations, $\xi_T(r)$ and $\eta_T(r)$ are the radial parts of
the  radial and the transverse component of the tidal displacement
with respect to the local coordinate basis $\partial/\partial r$,
$\partial/\partial \theta$, $\partial/\partial \phi$, and $\Psi_T(r)$
is the sum of the radial parts of the tide-generating potential and
the Eulerian perturbation of the star's gravitational
potential. Furthermore, $\rho$ is the density, $g$ the gravity, $c^2$
the square of the isentropic sound velocity, and $N^2$ the square of
the Brunt-V\"{a}is\"{a}l\"{a} frequency.  

The solutions of the system of Eqs.\ (\ref{dyn:1}) -- (\ref{dyn:3})
must satisfy boundary conditions at the star's centre and at the
star's surface. At $r=0$, the radial component of the tidal
displacement must remain finite. At $r=R_1$, the Lagrangian
perturbation of the pressure must vanish, and the continuity of the
gravitational potential and its gradient requires that
\begin{eqnarray}
\lefteqn{\left( {{d\Psi_T} \over {dr}} \right)_{R_1}
  + {{\ell+1} \over R_1}\, \left(\Psi_T\right)_{R_1}
  + \left( 4\, \pi\, G\, \rho\, \xi_T \right)_{R_1} } \nonumber \\
 & & = - \varepsilon_T (2\, \ell + 1) {{G\, M_1} \over R_1^2}\,
  c_{\ell,m,k}. \label{dyn:4}
\end{eqnarray}
Because of the non-homogeneous term in the right-hand member of this
boundary condition, the solutions of the system of Eqs.\ (\ref{dyn:1})
-- (\ref{dyn:3}) are proportional to the product $\varepsilon_T\,
c_{\ell,m,k}$. 

In the case of static tides, the forcing angular frequency $\sigma_T$
is equal to zero and the radial component of the tidal displacement is
determined by the homogeneous second-order differential equation
\begin{equation}
{{d^2 \xi_{T;0}} \over {dr^2}} 
  + 2\! \left( {1 \over g}\, {{dg} \over {dr}}
  + {1 \over r} \right)\! {{d\xi_{T;0}} \over {dr}} 
  - {{\ell(\ell+1) - 2} \over r^2}\, \xi_{T;0} = 0.
  \label{stat:1}
\end{equation}
This equation is equivalent to the equation of Clairaut which is usually
derived within the framework of the theory of equilibrium tides in the
limiting case of an infinite orbital period \citep{Ste1939}.

The solution of Eq.\ (\ref{stat:1}) must remain finite at $r=0$ and
satisfy the boundary condition 
\begin{equation}
\left( {{d \xi_{T;0}} \over {dr}} \right)_{R_1} +
  {{\ell-1} \over R_1}\, \left(\xi_{T;0}\right)_{R_1}
  = \varepsilon_T\, (2\, \ell + 1)\, c_{\ell,0,0}.
  \label{stat:2}
\end{equation}
 
For the remainder of the paper, we restrict the expansion of the
tide-generating potential given by Eq.~(\ref{pot}) to the terms
associated with $\ell=2$, which are dominant.

\section{Apsidal motion due to the tidal deformation of a component of
  a close binary}

The rate of secular apsidal motion due to the tidal distortion of the
components of a close binary is usually determined by means of the 
classical formula established in the assumption that the orbital
period is long in comparison to the periods of the free oscillation
modes of the component stars \citep{Cow1938,Ste1939}. At each instant,
the tide generated by 
the companion can then be approximated by a linear combination of 
three static tides (Paper~I). The resulting formula takes the form
\begin{equation}
\left( {{d \varpi} \over {dt}} \right)_{\rm classical}
  = \left({R_1\over a}\right)^5\, {M_2\over M_1}\,
  {{2\, \pi} \over P_{\rm orb}}\,
  k_2\, 15\, f\left({e^2}\right),  \label{strn:1}
\end{equation}
where $\varpi$ is the longitude of the periastron, and
\begin{equation}
f\left(e^2\right) = \left({1-e^2}\right)^{-5}\,
  \left({1 + {3\over 2}\, e^2 + {1\over 8}\, e^4 }\right). 
  \label{strn:2}
\end{equation}

The constant $k_2$ is known as the apsidal-motion constant and depends
on the mass concentration of the tidally distorted star. The constant is
determined as 
\begin{equation}
k_2 = {{\displaystyle 3 -  
  \left[ {{d\, \ln (\xi_{T;0}/r)} \over {d\, \ln r}} \right]_{R_1}}
  \over {\displaystyle 4 + 2\,  
  \left[ {{d\, \ln (\xi_{T;0}/r)} \over {d\, \ln r}} \right]_{R_1}}}.
  \label{k2}
\end{equation}
Because of the logarithmic derivatives in the numerator and the
denominator of this definition, the apsidal-motion constant is independent of
the product $\varepsilon_T\, c_{2,0,0}$ appearing in the boundary
condition given by Eq.\ (\ref{stat:2}). 

Within the framework of the theory of dynamic tides, the rate of
secular apsidal motion is determined by adding the contributions to
the apsidal motion stemming from the various partial static and
dynamic tides which are generated by the individual terms in the
expansion of the tide-generating potential given by
Eq.~(\ref{pot}). The resulting rate of secular change of the longitude
of the periastron is given by 
\begin{eqnarray}
\lefteqn{\left( {{d \varpi} \over {dt}} \right)_{\rm dyn} =
  \left({R_1 \over a}\right)^5 {M_2 \over M_1}\,
  {{2\, \pi} \over P_{\rm orb}}\, \bigg[
  2\, k_2\, G_{2,0,0} + 4\, \sum_{k=1}^\infty }
  \nonumber \\
 & & \left( F_{2,0,k}\, G_{2,0,k} + F_{2,2,k}\,
  G_{2,2,k} + F_{2,-2,k}\, G_{2,-2,k} \right) \bigg], 
  \label{aps:1}
\end{eqnarray}
where the $G_{2,m,k}$ are functions of the orbital eccentricity defined as
\begin{eqnarray}
\lefteqn{ G_{2,m,k}(e) = {1 \over {e\, \left( 1 - e^2 \right)^2}}\, 
  c_{2,m,k}\, P_2^{|m|}(0) } \nonumber \\
 & & {1\over \pi}\, \biggl[ 3 \int_0^\pi (1 + e\, \cos v)^2\,
  \cos (m\, v + k\, M)\, \cos v\, dv  \nonumber  \\
 & & - m \int_0^\pi (1 + e \cos v)\, (2 + e \cos v)  \nonumber \\
 & &  \sin (m\, v + k\, M)\, \sin v\, dv \biggr] \label{Glmk}
\end{eqnarray}
\citep{SWV1998}.

The coefficients $F_{2,m,k}$ render the star's responses to the
various forcing angular frequencies $\sigma_T$ and are related to the
total perturbation of the gravitational potential at the star's
surface as 
\begin{equation}
2\, F_{2,m,k} = - \left[ {R_1\over {G\, M_1}}
{{\Psi_T \left( R_1 \right)} \over
  {\varepsilon_T\, c_{2,m,k}}} + 1 \right].  \label{Flmk:1}
\end{equation}
In analogy with the apsidal-motion constant $k_2$, the coefficients
$F_{2,m,k}$ are independent of the product $\varepsilon_T\,
c_{2,m,k}$ which appears in the boundary condition given by Eq.\ (\ref{dyn:4}).

In the limiting case of long orbital and rotational periods, the
classical 
formula for the rate of secular apsidal motion agrees with the formula
established within the framework of the theory of dynamic tides up to
high orbital eccentricities \citep{SWV1998}. The agreement rests on
the property that, in this limiting case, all forcing angular
frequencies $\sigma_T$ tend to zero, so that the coefficients
$F_{2,m,k}$, with $m=-2,0,2$ and $k =1,2,3, \ldots$, tend to the 
apsidal-motion constant $k_2$. 

For binaries with shorter orbital periods, deviations arise due to the 
increasing role of the stellar compressibility at higher forcing
frequencies and due to resonances of dynamic tides with lower-order
$g^+$-modes of the component stars (Paper~I). The extent of the
deviations can be 
expected to depend on the evolutionary stage of the star due to the
increase of the stellar radius and the redistribution of the mass as 
the star evolves on the main sequence.

\section{Deviations of the classical apsidal-motion formula for
  binaries with an evolved main-sequence star}  

\subsection{The models}

\label{mod}

\begin{figure}
\resizebox{\hsize}{!}{\includegraphics{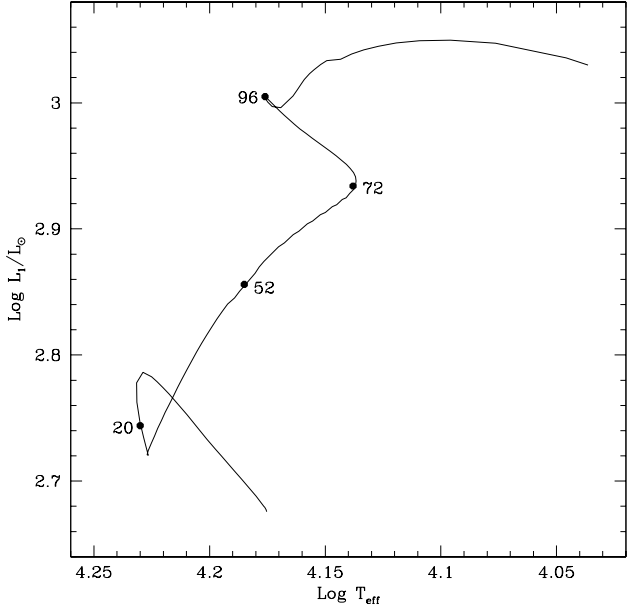}}
\caption{Evolutionary track of the $5\, M_\odot$ star with initial
  chemical composition $X=0.70$ and $Z=0.02$. The labels correspond to the 
  numbers of the models considered in our investigation.}  
\label{hrM5}
\end{figure}

We determined the deviations of the classical formula for the rate of
secular apsidal motion in binaries with shorter orbital periods for a
sequence of models of a $5\,M_\odot$ star with initial chemical
composition $X=0.70$ and $Z=0.02$.  The evolutionary track of the star
in the Hertzsprung-Russell diagram is displayed in
Fig.~\ref{hrM5}. The labels in the figure correspond to the numbers of
the models considered in our investigation, which range from the last
stages of the ${\rm C}^{12} \rightarrow {\rm N}^{14}$ reaction (model
20) to the phase where hydrogen is exhausted in the core (model~96).  

\begin{figure}
\resizebox{\hsize}{!}{\includegraphics{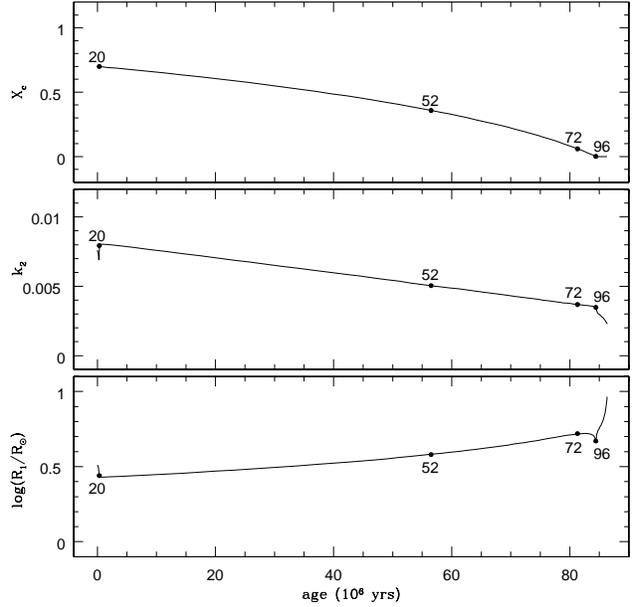}}
\caption{The variations of the central hydrogen abundance $X_c$, the
  apsidal-motion constant $k_2$, and the radius $R_1$ as a function of
  the age of the $5\, M_\odot$ star.}  
\label{evolM5}
\end{figure}

The variation of the central hydrogen abundance $X_c$, the
apsidal-motion constant $k_2$, and the radius $R_1$ are displayed in
Fig.~\ref{evolM5} as a function of the age of the star. The central 
hydrogen abundance $X_c$ decreases as the star evolves along the main
sequence due to the conversion of hydrogen into helium via the CNO cycle. 
In addition, the conversion of hydrogen into helium increases the central
condensation of the star, so that the value of the apsidal-motion constant $k_2$ decreases. 
At the end of the main sequence, when the star has reached an age of
$8.4 \times 10^7$ years, the central hydrogen abundance has dropped to
0.001 and the apsidal-motion constant has decreased by a factor of two. The 
radius of the star expands during the major part of the
evolution on the main sequence. It reaches a maximum value for
model~72 and subsequently decreases again during the final
stages of core-hydrogen burning. 
The ages, the radii, the central hydrogen abundances, and the
apsidal-motion constants of the models used in our investigation are
listed in Table~\ref{propM5}.  

\begin{table}
\caption{Properties of the models of the $5\, M_\odot$ star considered 
in this investigation.  \label{propM5}}  
\begin{tabular}{cccccccc}
\hline
 Model Nr.\ & age (yrs) & $R_1\, ({\rm km})$ & $X_c$ & $k_2$ \\
\hline \vspace{-0.33cm} \\
20 & $3.03 \; 10^5$ & $1.93 \; 10^6$ & 0.699 & 0.00741 \\
52 & $5.65 \; 10^7$ & $2.66 \; 10^6$ & 0.358 & 0.00505 \\
72 & $8.13 \; 10^7$ & $3.63 \; 10^6$ & 0.060 & 0.00369 \\
96 & $8.44 \; 10^7$ & $3.29 \; 10^6$ & 0.001 & 0.00348 \\
\hline
\end{tabular}
\end{table}

Due to the growth of the radius with the age of the star, care needs
to be taken that only orbital periods are considered for which the
radius of the star remains smaller than the radius of its Roche lobe. 
For binaries with
circular orbits, the star's Roche-lobe radius $R_{{\rm L},1}$ can be
approximated by means of Eggleton's (\citeyear{Egg1983}) fitting
formula 
\begin{equation}
{R_{{\rm L},1} \over a} = {{0.49\, q^{-2/3}} \over {0.6\, q^{-2/3}
  + \ln \left( 1 + q^{-1/3} \right)}},  \label{RL}
\end{equation}
where $q = M_2/M_1$. The relative error of the formula is smaller than
2\% for all values of the mass ratio in the range of $0<q<\infty$.  

\begin{figure}
\resizebox{\hsize}{!}{\includegraphics{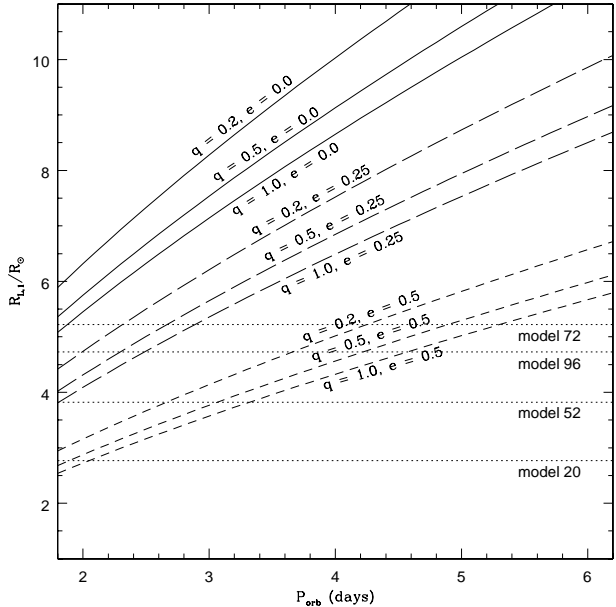}}
\caption{Variation of the Roche-lobe radius of a $5\,M_\odot$ star in
  the case of the orbital eccentricities $e=0.0, 0.25, 0.5$ and the
  mass ratios $q=0.2, 0.5, 1.0$. The dotted horizontal
  lines represent the radius of the star for models 20, 52, 72, and 96.}   
\label{frl1}
\end{figure}

The variation of the Roche-lobe radius of a $5\,M_\odot$ star,
expressed in solar radii, is presented in Fig.~\ref{frl1} as a
function of the orbital period for the mass ratios $q=0.2, 0.5,
1.0$. The solid lines in the figure correspond to the Roche-lobe radii
for binaries with a circular orbit. The dotted horizontal 
lines represent the stellar radii at the evolutionary stages
represented by the models 20, 52, 72, and 96.

In the case of a binary with an eccentric orbit, the Roche-lobe radius
can be expected to be smallest when the companion is located in the
periastron of its relative orbit. Therefore, we also determined the
variations of the Roche-lobe radius 
for binaries with an eccentric orbit by substituting the
periastron distance $r_p=a(1-e)$ for the semi-major axis $a$ in Eq.\
(\ref{RL}). The resulting Roche-lobe radii are represented by the
long- and the short-dashed lines in Fig.~\ref{frl1} 
for the orbital eccentricities $e=0.25$ and $e=0.5$, respectively.  

The Roche-lobe radii increase monotonically with increasing values of
the orbital
period and are larger for smaller values of the mass ratio. The
restrictions on the orbital period imposed by the requirement that the
star fit within its Roche-lobe are most stringent for higher orbital
eccentricities. In the case of model~72, e.g., the orbital period must
be larger than 3~days when the eccentricity takes the value $e=0.25$
and 
larger than 5.5~days when the eccentricity takes the value $e=0.5$.

\subsection{Deviations of the classical formula}

An appropriate quantity to examine the deviations between the rate of
secular apsidal motion given by the classical formula and the
corresponding 
rate determined within the framework of the theory of dynamic tides is
the relative difference
\begin{equation}
\Delta = {{\displaystyle \left( {d \varpi/dt} \right)_{\rm classical}
  - \left( {d\varpi/dt} \right)_{\rm dyn}} \over {\displaystyle
  \left( {d\varpi/dt} \right)_{\rm  dyn}}}.  \label{Delta}
\end{equation}
This relative difference is independent of the product 
\[
\left(R_1/a\right)^5 \left(M_2/M_1\right) 
  \left(2\, \pi/ P_{\rm orb}\right), 
\]
so that it depends on the orbital period only through the
coefficients $F_{2,m,k}$. 

\begin{figure*}
\resizebox{8.5cm}{!}{\includegraphics{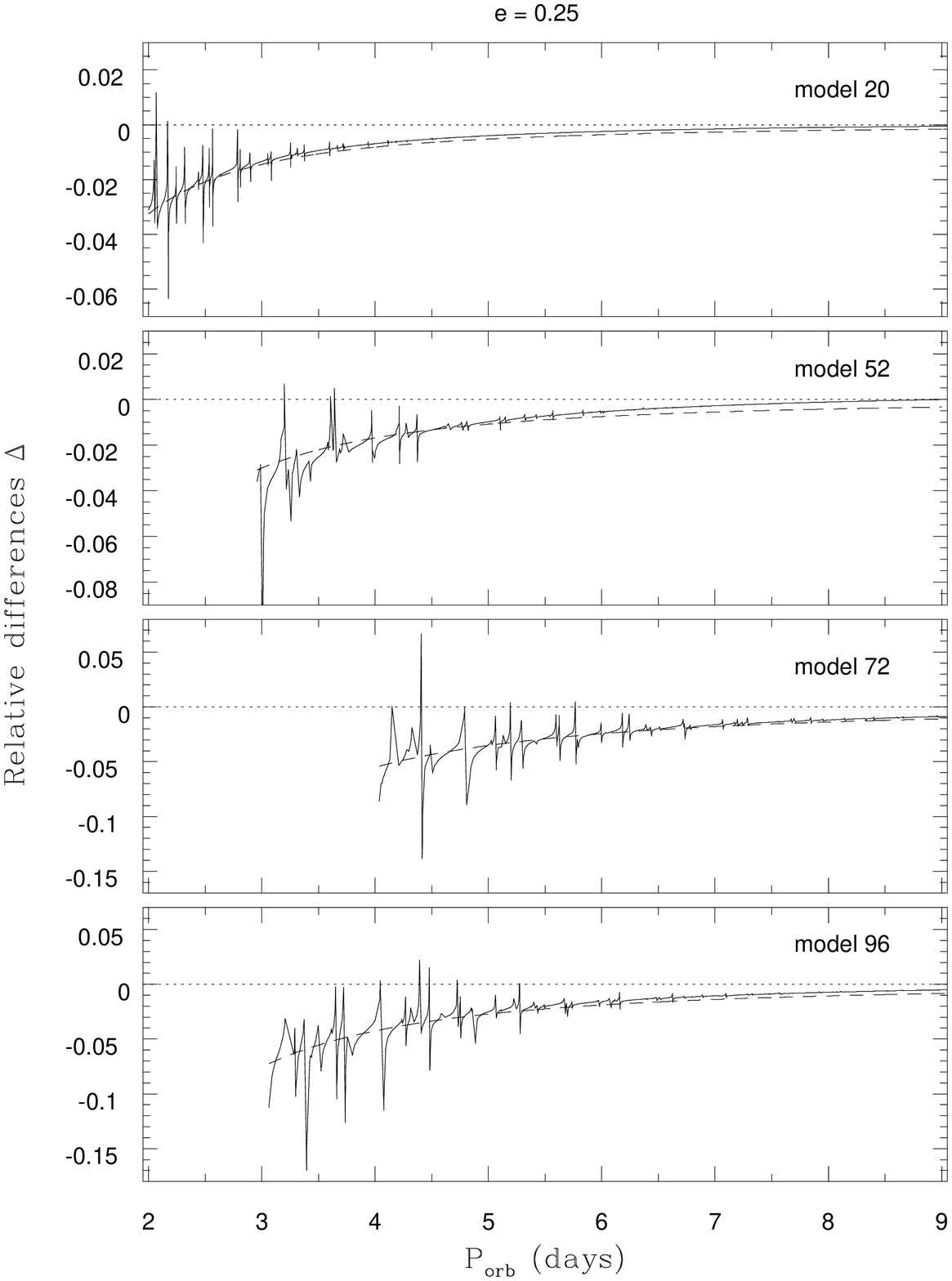}}
\resizebox{8.5cm}{!}{\includegraphics{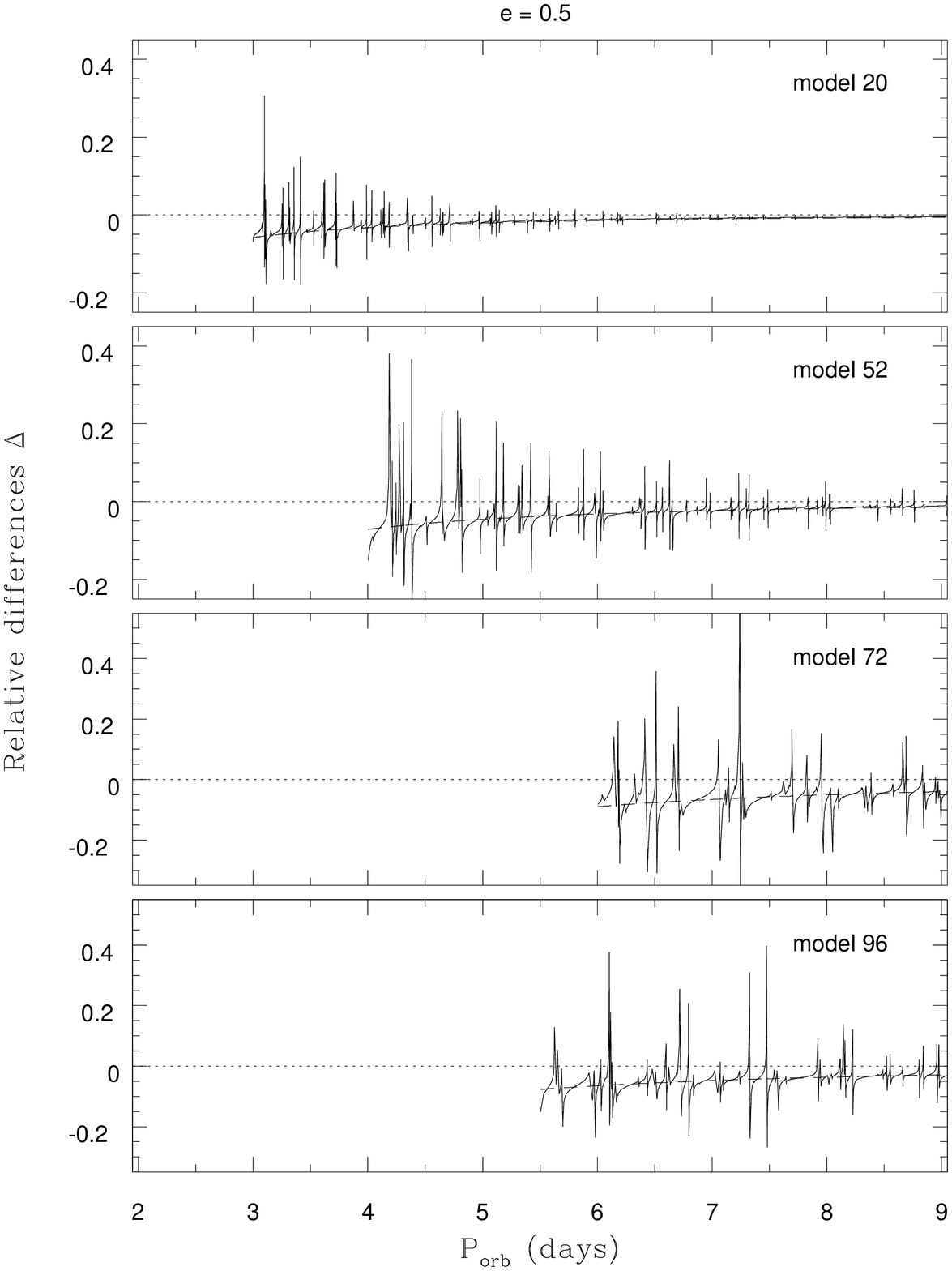}}
\caption{The relative differences $\Delta$ for models 20, 52, 72, and 96 of
  the $5\, M_\odot$ star in the case of the orbital eccentricities $e=0.25$ 
  (left-hand panel) and $e=0.5$ (right-hand panel). The solid lines 
  represent the
  relative differences in the presence of resonances of dynamic tides
  with free oscillation modes $g^+$. The dashed lines correspond to
  the relative differences in the absence of resonances.}   
\label{aps_ev}
\end{figure*}

We determined the relative differences $\Delta$ as a function of the
orbital period for the evolutionary models 20, 52, 72, and 96 of the
$5\,M_\odot$ star, 
and the orbital eccentricities $e=0.25$ and 
$e=0.5$. The shortest orbital periods considered for each model are
chosen such that the radius of the star is sufficiently small in
comparison to the radius of its Roche lobe (Sect.~\ref{mod}). For the
determination of the forcing angular frequencies $\sigma_T$ with
respect to the corotating frame of reference, we adopted the low
angular velocity of rotation $\Omega=0.01\, n$, so that the
companion's orbital period with respect to the corotating frame 
remains short.  In the case of a resonance of a dynamic tide
with a free oscillation mode, we limit the determination of $\Delta$
to orbital periods for which the relative difference between the
forcing angular frequency of the tide and the eigenfrequency of the
oscillation mode is larger than $0.1\, \varepsilon_T$
(see Paper~I). The resulting 
variations of the relative difference $\Delta$ are represented by the
solid lines in the left- and right-hand panel of Fig.~\ref{aps_ev}.  

The relative differences $\Delta$ are mostly {\em negative} so that
for binaries with shorter orbital periods the classical formula yields
somewhat too small values  
for the rate of secular apsidal motion, and thus somewhat too long
apsidal-motion periods. The peaks observed at shorter orbital periods
are caused by resonances of dynamic tides with free oscillation modes
$g^+$ of the tidally distorted star. These peaks are superposed on a
basic curve which represents the systematic deviations caused by the
increasing role of the stellar compressibility at shorter orbital
periods.  

The systematic relative differences $\Delta$ in the absence of
resonances can be approximated by a formula of the form 
\begin{equation}
\Delta_{\rm basic} = - \lambda\, T_{\rm orb}^{-2},
  \label{basic}
\end{equation}
where $\lambda$ is a constant. For the determination of $\lambda$ we
used a linear least-squares procedure described by \citet{Numrec} and
fitted Eq.\ (\ref{basic}) to the relative differences $\Delta$
displayed in Fig.~\ref{aps_ev}. The resulting values of $\lambda$ are
listed in  Table~\ref{lambda} in the supposition that the orbital
period is expressed in days. The associated basic curves determined by
means of Eq.\ (\ref{basic}) are represented by the dashed lines in
Fig.~\ref{aps_ev}. 

\begin{table}
\caption{The proportionality factor $\lambda$ for models 20, 52, 72, and
  96 of the $5\, M_\odot$ star and the orbital eccentricities $e=0.25$ and
  $e=0.5$.  \label{lambda}}  
\begin{tabular}{ccc}
\hline
\hspace{0.25cm} Model Nr.\ \hspace{0.25cm} & \hspace{0.25cm}
$\lambda_{e=0.25}$ \hspace{0.25cm} & \hspace{0.25cm} $\lambda_{e=0.5}$
\hspace{0.25cm} \\ 
\hline \vspace{-0.33cm} \\
20 & 0.13 & 0.52  \\
52 & 0.27 & 1.1  \\
72 & 0.88 & 3.2  \\
96 & 0.68 & 2.3  \\
\hline
\end{tabular}
\end{table}

The downwards slope of the basic curves is larger for a model
representing a more evolved stage of the $5\, M_\odot$ star. The slope
is largest for model 72, which is the model with the largest radius,
and decreases again during the final stages of core-hydrogen
burning. The underlying reason can be identified by expressing the
forcing angular frequencies $\sigma_T$ in units of the inverse of the
star's dynamical time scale:  
\begin{equation}
\sigma_T = \left( {{G\, M_1} \over R_1^3} \right)^{1/2} \sigma_T^\ast.
  \label{sigT}
\end{equation}
For a model with a constant mass, it then follows that the
dimensionless forcing angular frequencies $\sigma_T^\ast$ increase
with increasing values of the stellar radius. The systematic
deviations caused by the compressibility of the stellar fluid will
therefore be larger for a model with a larger radius. 
In the case of the orbital eccentricity $e=0.5$ and the orbital period
of 6 days, e.g., the relative differences $\Delta$ in the absence of
resonances increase from 1.4\% for model~20 to 8.9\% for model~72.  

In the case of a resonance of a dynamic tide with a free oscillation
mode, the relative differences $\Delta$ may deviate substantially from
the systematic trend represented by the basic curves. The largest
deviations occur for resonances between the lower-order harmonics in
the expansion of the tide-generating potential given by
Eq.~(\ref{pot}) and the 
lower-order $g^+$-modes of the tidally distorted star. For these
resonances the relative differences between the rates of secular
apsidal motion predicted by the classical formula and the rates
given by the theory of dynamic tides can amount to as much as a few
tens of percents, especially for higher orbital eccentricities.  
This conclusion is in agreement with the conclusions reached by
 \citet{PP1980} and \citet{Qua1996}.

For a given range of orbital periods and a fixed orbital eccentricity,
the deviations caused by the resonances increase as the star evolves
on the main sequence.
This behaviour is related to the growing radiative
envelope which causes the eigenfrequencies of a more evolved stellar
model to be smaller and closer to each other than those of a less evolved
stellar model, so that  resonances with lower-order
$g^+$-modes become increasingly important with increasing age of the
star. A similar increase in the effectiveness of tidal 
forcing during the late stages of core hydrogen burning was found by
\citet{SP1984} in an investigation on the spin-up of neutron stars and the
circularization of orbits in massive X-ray binaries.

\section{Concluding remarks}

In this paper, we have examined the influence of stellar evolution on
the deviations between the rates of secular apsidal motion predicted
by the  classical formula due to \citet{Cow1938} and
\citet{Ste1939}, and the rates of secular apsidal motion determined 
in the framework of the theory of dynamic tides. The deviations arise
for binaries with shorter orbital periods due to the growing 
role of the stellar compressibility at higher forcing frequencies and
due to resonances of dynamic tides with the free oscillation modes of the
component stars \citep{SW2001}.  

The extent of the deviations caused by the compressibility of the
stellar fluid depends on the evolutionary stage of the star primarily
through the evolution of the radius and the associated change of the
star's dynamical time scale. As the star evolves on the main sequence,
the radius and the dynamical time scale increase so that the orbital
period effectively becomes shorter in comparison to the star's
dynamical time scale. The deviations due to the compressibility of the
stellar fluid are therefore larger for a model with a larger radius.  

Besides varying the systematic deviations caused by the
compressibility of the stellar fluid, the evolution of the star also
affects the deviations caused by the resonances of dynamic tides with
free oscillation modes of the component stars. The latter
deviations are superposed upon the deviations caused by the stellar
compressibility and can amount to as much as ten percent and more,
especially for higher orbital eccentricities. In the case of the $5\,
M_\odot$ star considered, the effects of the resonances are larger for
a model near the end of core-hydrogen burning than for a model near
the zero-age main sequence.

\begin{acknowledgements}
Bart Willems acknowledges the support of PPARC grant PPA/G/S/1999/00127. 
\end{acknowledgements}

\aareferences

\end{document}

%% file: rrmacros2e.tex
\newcount\longrefs
\def\adaa{\ifnum\longrefs=1 {Advances in Astronomy and Astrophysics}\else 
                           {Adv. Astron.\ Astrophys.}\fi}
\def\aap{\ifnum\longrefs=1 {Astron.\ Astrophys.}\else 
                           {A\hbox{\rm \&}A}\fi}
\def\aapr{\ifnum\longrefs=1 {Astron.\ Astrophys.\ Rev.}\else 
                            {A\hbox{\rm \&}AR}\fi}
\def\aaps{\ifnum\longrefs=1 {Astron.\ Astrophys.\ Suppl.}\else 
                            {A\hbox{\rm \&}AS}\fi}
\def\aj{\ifnum\longrefs=1 {Astron.\ J.}\else 
                          {AJ}\fi} 
\def\ao{\ifnum\longrefs=1 {Applied Optics}\else 
                           {Appl.\ Opt.}\fi} 
\def\aspcs{\ifnum\longrefs=1 {Astron.\ Soc.\ Pacific Conf. Series}\else 
                           {ASP Conf.\ Ser.}\fi} 
\def\apj{\ifnum\longrefs=1 {Astrophys.\ J.}\else 
                           {ApJ}\fi} 
\def\apjl{\ifnum\longrefs=1 {Astrophys.\ J. Lett.}\else 
                            {ApJ}\fi} 
\def\aplett{\ifnum\longrefs=1 {Astrophys.\ J. Lett.}\else 
                            {ApJ}\fi} 
\def\apjs{\ifnum\longrefs=1 {Astrophys.\ J. Suppl.}\else 
                            {ApJS}\fi}
\def\apss{\ifnum\longrefs=1 {Astrophys.\ and Space Science}\else 
                            {Ap\&SS}\fi}
\def\araa{\ifnum\longrefs=1 {Ann.\ Rev.\ Astron.\ Astrophys.}\else 
                            {ARA\hbox{\rm \&}A}\fi}
\def\azh{\ifnum\longrefs=1 {Astronomicheskii Zhurnal}\else 
                            {Astron.\ Zhur.}\fi}
\def\baas{\ifnum\longrefs=1 {Bull.\ Am.\ Astron.\ Soc.}\else 
                            {BAAS}\fi}
\def\bain{\ifnum\longrefs=1 {Bull.\ Astronom.\ Institutes Netherlands}\else
                            {Bull.\ Astr.\ Inst.\ Neth.}\fi}
\def\gca{\ifnum\longrefs=1 {Geochim.\ Cosmochim.\ Acta}\else 
                           {Geochim.\ Cosmochim.\ Acta}\fi}
\def\geo{\ifnum\longrefs=1 {Geophysical Journal}\else 
                           {Geophys.\ J.}\fi}
\def\grl{\ifnum\longrefs=1 {Geophys.\ Res.\ Lett.}\else 
                           {Geoph.\ Res.\ Lett.}\fi}
\def\iaucirc{\ifnum\longrefs=1 {IAU Circulars}\else 
                          {IAU Circ.}\fi}
\def\ip{\ifnum\longrefs=1 {in press}\else 
                          {in press}\fi}
\def\jgr{\ifnum\longrefs=1 {J.\ Geophys.\ Res.}\else 
                           {J.\ Geophys.\ Res.}\fi}  
\def\jrasc{\ifnum\longrefs=1 {J.\ Royal Astron.\ Soc.\ Canada}\else 
                           {JRAS Can.}\fi}  
\def\mnras{\ifnum\longrefs=1 {Mon.\ Not.\ Roy.\ Astron.\ Soc.}\else 
                             {MNRAS}\fi} 
\def\nat{\ifnum\longrefs=1 {Nature}\else 
                           {Nat}\fi}
\def\pasj{\ifnum\longrefs=1 {Pub.\ Astron.\ Soc.\ Japan}\else 
                            {PASJ}\fi} 
\def\pasp{\ifnum\longrefs=1 {Pub.\ Astron.\ Soc.\ Pacific}\else 
                            {PASP}\fi} 
\def\physscr{\ifnum\longrefs=1 {Physica Scripta}\else 
                            {Phys.\ Scrip.}\fi} 
\def\planss{\ifnum\longrefs=1 {Planetary \& Space Science}\else 
                            {Plan. \& Space Sci.}\fi} 
\def\procspie{\ifnum\longrefs=1 {Proc.\ SPIE}\else 
                            {Proc.\ SPIE}\fi} 
\def\qjras{\ifnum\longrefs=1 {Quarterly J.\ Royal Astron.\ Soc.}\else 
                            {QJRAS}\fi} 
\def\sa{\ifnum\longrefs=1 {Soviet Astron..}\else 
                               {Sov.\ Astron.}\fi}
\def\skytel{\ifnum\longrefs=1 {Sky \& Telescope}\else 
                            {Sky \& Tel.}\fi} 
\def\solphys{\ifnum\longrefs=1 {Solar Phys.}\else 
                               {Solar Phys.}\fi}
\def\ssr{\ifnum\longrefs=1 {Space Science Rev.}\else 
                               {Space\ Sci.\ Rev.}\fi}

\def\bibfiles{/home/bwillems/latex/bibtex/bibliofile}   

\def\aareferences{\longrefs=0  \bibliographystyle{/home/bwillems/latex/bibtex/aabib}
             \bibliography{/home/bwillems/latex/bibtex/aajour,\bibfiles}}

\def\dutch{\def\refname{Referenties}\def\abstractname{Samenvatting}%
  \def\bibname{Bibliografie}\def\chaptername{Hoofdstuk}%
  \def\appendixname{Bijlage}\def\contentsname{Inhoudsopgave}%
  \def\listfigurename{Lijst van figuren}\def\listtablename{Lijst van tabellen}%
  \def\indexname{Index}\def\figurename{Figuur}\def\tablename{Tabel}%
  \def\partname{Deel}\def\enclname{Bijlage(n)}\def\ccname{Ter attentie van}%
  \def\headtoname{Aan}\def\headpagename{Pagina}%
  \def\today{\number\day\space\ifcase\month\or januari\or februari\or maart\or%
     april\or mei\or juni\or juli\or augustus\or september\or oktober\or%
     november\or december\fi \space\number\year}%
  \typeout{
              >>>>> use hlatex209 for Dutch hyphenation <<<<< 
         }}
\newcounter{onefig} \newcounter{fignumber}
\newcount\nocaptions \newcount\nofigures \newcount\figwidth
\newcount\viewgraphs
  \def\paper{}  \def\figlabel{} 
\long\def\nextfig#1{\setcounter{figure}{\value{fignumber}}
  \addtocounter{fignumber}{1}
  \ifnum \viewgraphs=1 \newpage \pagestyle{empty} \fi 
  \ifnum\value{onefig}=0 #1 \fi                 
  \ifnum\value{onefig}=\value{fignumber} #1 \fi}
\def\figwidths#1#2{\ifnum \nocaptions=1 #2mm \else #1mm \fi}  
\def\paper#1{}  
\long\def\plotfig#1#2{\ifnum \nofigures=1 \else #2 \fi}
\long\def\captiontext#1{\ifnum \nofigures=1 \raggedright \fi 
   \ifnum \nocaptions=1 \paper
     \ifnum \viewgraphs=0 
       \newline  \mbox{}\hrulefill\mbox{} \newline 
       \newline label:~\{\figlabel\} 
     \fi 
     \else \ifnum \nofigures=0 \fi 
   #1 \fi}

\newcount\panelwidth \newcount\panelheight 
\newcount\bxmin \newcount\bymin \newcount\bxmax \newcount\bymax
\newcount\tbxmin \newcount\tbymin
\newcount\tpanelwidth \newcount\tpanelheight \newcount\tpdif
\def\panelsize #1,#2;{\panelwidth=#1 \panelheight=#2}  
\def\setbb #1,#2;#3,#4;#5,#6;{
  \tbxmin=#1 \tbymin=#2    
  \bxmin=#3 \bymin=#4      
  \bxmax=#5 \bymax=#6}     
\def\barepanel #1{%
  \ifnum\panelheight=0 
    \tpdif=\bymax \advance\tpdif by -\bymin
    \multiply \tpdif by \panelwidth
    \tpanelheight=\tpdif
    \tpdif=\bxmax \advance\tpdif by -\bxmin
    \divide \tpanelheight by \tpdif
  \else \tpanelheight=\panelheight \fi
  \epsfig{file=#1,%
     bbllx=\bxmin bp,bblly=\bymin bp,bburx=\bxmax bp,bbury=\bymax bp,clip=,%
     width=\panelwidth mm,height=\tpanelheight mm}}
\def\labelypanel #1{
  \ifnum\panelheight=0 
    \tpdif=\bymax \advance\tpdif by -\bymin
    \multiply \tpdif by \panelwidth
    \tpanelheight=\tpdif
    \tpdif=\bxmax \advance\tpdif by -\bxmin
    \divide \tpanelheight by \tpdif
  \else \tpanelheight=\panelheight \fi
  \tpdif=\bxmax \advance\tpdif by -\tbxmin
  \tpanelwidth=\panelwidth \multiply \tpanelwidth by \tpdif
  \tpdif=\bxmax \advance\tpdif by -\bxmin
  \divide \tpanelwidth by \tpdif
  \epsfig{file=#1,%
    bbllx=\tbxmin bp,bblly=\bymin bp,bburx=\bxmax bp,bbury=\bymax bp,%
    clip=,width=\tpanelwidth mm,height=\tpanelheight mm}}
\def\labelxpanel #1{%
  \ifnum\panelheight=0 
    \tpdif=\bymax \advance\tpdif by -\bymin
    \multiply \tpdif by \panelwidth
    \tpanelheight=\tpdif
    \tpdif=\bxmax \advance\tpdif by -\bxmin
    \divide \tpanelheight by \tpdif
  \else \tpanelheight=\panelheight \fi
  \tpdif=\bymax \advance\tpdif by -\tbymin
  \multiply \tpanelheight by \tpdif
  \tpdif=\bymax \advance\tpdif by -\bymin
  \divide \tpanelheight by \tpdif
  \epsfig{file=#1,%
    bbllx=\bxmin bp,bblly=\tbymin bp,bburx=\bxmax bp,bbury=\bymax bp,%
    clip=,width=\panelwidth mm,height=\tpanelheight mm}}
\def\labelxypanel #1{%
  \ifnum\panelheight=0 
    \tpdif=\bymax \advance\tpdif by -\bymin
    \multiply \tpdif by \panelwidth
    \tpanelheight=\tpdif
    \tpdif=\bxmax \advance\tpdif by -\bxmin
    \divide \tpanelheight by \tpdif
  \else \tpanelheight=\panelheight \fi
  \tpdif=\bxmax \advance\tpdif by -\tbxmin
  \tpanelwidth=\panelwidth \multiply \tpanelwidth by \tpdif
  \tpdif=\bxmax \advance\tpdif by -\bxmin
  \divide \tpanelwidth by \tpdif 
  \tpdif=\bymax \advance\tpdif by -\tbymin 
  \multiply \tpanelheight by \tpdif
  \tpdif=\bymax \advance\tpdif by -\bymin
  \divide \tpanelheight by \tpdif
  \epsfig{file=#1,%
    bbllx=\tbxmin bp,bblly=\tbymin bp,bburx=\bxmax bp,bbury=\bymax bp,%
    clip=,width=\tpanelwidth mm,height=\tpanelheight mm}}

\def\CC{\par \vspace*{-2ex} \footnotesize \baselineskip=8pt \begin{verbatim}}

\long\def\startignore #1\stopignore{}   

\def\setlistparams{         
  \topsep=0.7ex                 
  \itemsep=0.7ex                
  \leftmargini=3ex}             
\setlistparams                  

\newcounter{alistindex}       

\newcounter{romenumnr}

\newlength{\minipagewidth}

\newsavebox{\boxcontent}
\newcommand{\ovalhead}[1]{
  \unitlength=1cm
  \sbox{\boxcontent}{\mbox{~~{#1}~~}}
  \begin{center}
    \ifdim\wd\boxcontent>6ex 
    \ifdim\wd\boxcontent<8cm 
    \begin{picture}(8,3) \thicklines     
      \put(4.0,0.8){\oval(8,1.6)} 
      \put(0.0,0.7){\parbox{8cm}{
         \begin{center} \usebox{\boxcontent} \end{center}}}
    \end{picture}
    \else \ifdim\wd\boxcontent<12cm 
    \begin{picture}(12,3) \thicklines     
        \put(6.0,0.8){\oval(12,1.6)} 
        \put(0.0,0.7){\parbox{12cm}{
           \begin{center} \usebox{\boxcontent} \end{center}}}
    \end{picture}
    \else
    \begin{picture}(16,3) \thicklines     
        \put(8.0,0.8){\oval(16,1.6)} 
        \put(0.0,0.7){\parbox{16cm}{
           \begin{center} \usebox{\boxcontent} \end{center}}}
    \end{picture}
    \fi \fi \fi
  \end{center}} 

\newcounter{headnr}            
\newcounter{subheadnr}[headnr]
\newcounter{subsubheadnr}[subheadnr]
\def\head #1\par{
  \stepcounter{headnr}                          
  \vspace{2ex} \noindent                        
  {\bf \theheadnr~~~~#1}\\[1ex] \noindent}      
\def\subhead #1\par{  
  \stepcounter{subheadnr}
  \vspace{1.3ex} \noindent
  {\bf \theheadnr.\arabic{subheadnr}~~~#1}\\[0.3ex] \noindent}
\def\subsubhead #1\par{
  \stepcounter{subsubheadnr}
  \vspace{1.0ex} \noindent
  {\bf \theheadnr.\arabic{subheadnr}.\arabic{subsubheadnr}~~~#1}\\ \noindent}

\font\dropfont= cmr12 scaled \magstep5
\def\dropcap#1#2{{\noindent
    \setbox0\hbox{\dropfont #1}\setbox1\hbox{#2}\setbox2\hbox{(}%
    \count0=\ht0\advance\count0 by\dp0\count1\baselineskip
    \advance\count0 by-\ht1\advance\count0by\ht2
    \dimen1=.5ex\advance\count0by\dimen1\divide\count0 by\count1
    \advance\count0 by1\dimen0\wd0
    \advance\dimen0 by.25em\dimen1=\ht0\advance\dimen1 by-\ht1
    \global\hangindent\dimen0\global\hangafter-\count0
    \hskip-\dimen0\setbox0\hbox to\dimen0{\raise-\dimen1\box0\hss}%
    \dp0=0in\ht0=0in\box0}#2}

\def\level #1 #2#3#4{$#1 \: ^{#2} \mbox{#3} ^{#4}$}   

\def\mathstacksym#1#2#3#4#5{\def#1{\mathrel{\hbox to 0pt{\lower 
    #5\hbox{#3}\hss} \raise #4\hbox{#2}}}}

\mathstacksym\lta{$<$}{$\sim$}{1.5pt}{3.5pt} 
\mathstacksym\gta{$>$}{$\sim$}{1.5pt}{3.5pt} 
\mathstacksym\lrarrow{$\leftarrow$}{$\rightarrow$}{2pt}{1pt} 
\mathstacksym\lessgreat{$>$}{$<$}{3pt}{3pt} 
